\documentclass[runningheads]{llncs}
\usepackage{amssymb}
\usepackage{amsmath}
 
\begin {document}

\newcommand{\bL}{{\mathbf L}}

\newcommand{\bU}{{\mathbf U}}
\newcommand{\bA}{{\mathbf A}}
\newcommand{\bB}{{\mathbf B}}
\newcommand{\bC}{{\mathbf C}}
\newcommand{\bD}{{\mathbf D}}

\newcommand{\Dmtr}[4]%
{\left|\begin{array}{rr}#1 & #2\\#3 & #4\end{array}\right|}
\def\cA{{\bf \mathcal A}}
\def\cB{{\bf \mathcal B}}
\def\cC{{\bf \mathcal C}}
\def\cD{{\bf \mathcal D}}

\newcommand{\diag}{{\mathrm{diag}}}

\def\al{{\alpha}}
\def\be{\beta}
\def\de{\delta}

\def\ga{\gamma}

\def\cA{{\bf \mathcal A}}
\def\cU{{\bf \mathcal U}}
\def\cL{{\bf \mathcal L}}

\title{Triangular Decomposition of Matrices in a Domain
\thanks{ Preprint of the paper: G.Malaschonok, A.Scherbinin. Triangular Decomposition of Matrices in a Domain. 
Computer Algebra in Scientific Computing. LNCS 9301, Springer, Switzerland, 2015, P.290-304.}
}
\titlerunning{Triangular Decomposition of Matrices}
\author{Gennadi Malaschonok and Anton Scherbinin}
\authorrunning{G.Malaschonok, A.Scherbinin}  
\institute{Tambov State University, \\  Internatsionalnaya 33,
392622 Tambov, Russia \\
\email{malaschonok@gmail.com}}

\maketitle

\begin{abstract}
Deterministic recursive algorithms for the computation of matrix triangular decompositions with permutations 
like LU and  Bruhat decomposition are presented for the case of commutative domains.
This decomposition can be considered as a generalization of LU and Bruhat decompositions, 
because they both may be easily obtained from this triangular decomposition.
Algorithms have the same complexity as the algorithm of matrix multiplication.
\end{abstract}
%
\section{Introduction} 
Traditionally, decomposition of   matrices in a product of several triangular matrices and permutation 
matrices is investigated for matrices over fields.

Active use of functional and polynomial matrices in computer algebra systems has 
generated interest to matrix decomposition in commutative domains.

Matrix decomposition of the form  $A=VwU$ is called the Bruhat decomposition, if
$V$ and $U$ are nonsingular upper triangular matrices and $w$ is a
matrix of permutation. The generalized Bruhat decomposition was introduced  by Dima Grigoriev \cite{08},\cite{09}.

Matrix decomposition of the form  $A=LU$ or $A=PLUQ$ is called LU-decomposition of the matrix $A$, if
$L$ and $U$ are lower and upper triangular matrices and $P$, $Q$ is the matrices of permutations.

In general, for an arbitrary matrix, we can not get the Bruhat decomposition based on the expansion LU. And vice versa.

In this paper we suggest a new form of triangular decomposition. We call it $LDU$-decomposition.
Here $L$ and $U$ are lower and upper triangular matrices,  $D=PdQ$, where $d$ -- a diagonal matrix, $P$ and $Q$ -- permutation matrices.

This decomposition can be considered as a generalization of LU and Bruhat decompositions. 
Since they both may be easily obtained from $LDU$-decomposition.

We describe and prove the deterministic recursive algorithm for computation of such triangular decomposition.
This algorithm have the same complexity as the algorithm of matrix multiplication (proof see in \cite{18}).
In \cite{18} there ware described particular cases of this algorithm, when each of the main corner minor of a 
matrix $A$ up to the rank is not equal zero.

Now we present the complete algorithm for the general case and give its proof.

\section{Preliminary. Triangular Decomposition in Domain}

Let $R$ be a commutative domain, $A=(a_{i,j}) \in R^{n \times m}$ be a matrix of size $n\times m$, 
$\alpha^k_{i,j}$ be $k\times k$ minor of matrix $A$ which disposed in the rows $1,2, \ldots ,k-1,i$ and columns   $1,2, \ldots ,k-1,j$. We denote $\alpha^0=1$ and $\alpha^k = \alpha^k_{k,k}$. And we use the notation $\de_{ij}$ for Kronecker delta.
\smallskip

Let $k$ and $s$ be  integers in the interval $0\leq k<s\leq n$, $\cA_{s,p}^k=(\al^{k+1}_{i,j})$ be the matrix of minors 
with size $(s-k)\times (p-k)$ which has elements $\al^{k+1}_{i,j}$, $i=k+1, \ldots ,s-1,s$,  $j=k+1, \ldots ,p-1,p$  and  $\cA_n^0=(\al^{1}_{i,j})=A$. We denote $\cA_{s,s}^k=\cA_{s}^k$.

We shall use the following identity (see \cite{10}, \cite{12}):
\begin{theorem}[Sylvester determinant identity]

\noindent Let $n=m$, $k$ and $s$ be the integers in the interval $0\leq
k<s\leq n$. Then it is true that
\begin{equation}
\label{eq_1}
\det (\cA_s^k)=\al^s (\al^k)^{s-k-1}.
\end{equation}
\end{theorem}
\begin{theorem}[LDU decomposition of the minors matrix]

\noindent Let $A=(a_{i,j}) \in R^{n \times m}$ be the matrix of rank $r$, $\alpha^i\neq 0$ for $i=k,k+1, \ldots ,r$, then the matrix of minors $\cA_{s,p}^k$, $k<s\le n$, $k<p\le m$ 
is equal to the following product of three matrices:

\begin{equation}
\label{eq_2}
\cA_{s,p}^k = L^k_sD^k_{s,p}U^k_p  =(a^j_{i,j})  (\de_{ij}\alpha^k(\alpha^{i-1} \alpha^{i} )^{-1}) (a^i_{i,j}).
\end{equation}

\noindent The matrix $ L^k_s=(a^j_{i,j})$, $i=k+1 \ldots s$, $j=k+1 \ldots r$, is a low triangular matrix of size $(s-k)\times (r-k)$,
 the matrix $U^k_p=(a^i_{i,j})$, $i=k+1 \ldots r$, $j=k+1 \ldots p$, is an upper triangular matrix of size $(r-k)\times (p-k)$ and
$D^k_{s,p} = (\de_{ij}\alpha^k(\alpha^{i-1} \alpha^{i} )^{-1})$, $i=k+1 \ldots r$, $j=k+1 \ldots r$, is a diagonal matrix of size $(r-k)\times (r-k)$.
\end{theorem}

\begin{proof} 
 See proof in \cite{18}.
\end{proof} 

\begin{remark} 
We can add the unit blocks to the matrices  $ L^k_s$ and $U^k_p$ and we obtain nonsingular square matrices of size $(s-k)$
and $(p-k)$ respectively. We can add zero elements to the matrix $D^k_{s,p}$ and obtain the diagonal matrix of size
$(s-k)\times (p-k)$. As a result we obtain decomposition $\cA_{s,p}^k = L^k_sD^k_{s,p}U^k_p$ with invertible square matrices $L^k_s$ and $U^k_p$ 
\end{remark}

\begin{corollary}[LDU decomposition of matrix $A$ in domain]
\par\noindent

Let $A=(a_{i,j}) \in R^{n \times m}$, be the matrix of rank $r$,  $\alpha^i\neq 0$ for
$i=1,2, \ldots ,r$, then matrix  $A$ is equal to the following product of three matrices:
\begin{equation}
\label{eq_3}
 A = L^0_n D^0_{n,m} U^0_m = (a^j_{i,j})  (\de_{ij} (\alpha^{i-1} \alpha^{i} )^{-1}) (a^i_{i,j}).
\end{equation}
\end{corollary}

\begin{corollary}[Bruhat decomposition]

\par\noindent 

Let $A=LDU$ be LDU-decomposition of a $n\times n$ matrix $A$ with rank $r$, $S$
be the "flipped" identity matrix, then
$V=SLS$ and $U$ are  upper triangular matrices  
and
\begin{equation}
\label{eq_4}
SA=V (SD) U
\end{equation}

is the Bruhat decomposition of the matrix $SA$.
\end{corollary}

\begin{corollary}[Cases of zero blocks]
Let $\cA_{s,p}^k=\left( \begin{array}{cc}\bA & \bB \\ \bC & \bD \end{array}\right)$,
$S_2= \left(\begin{array}{cc} 0 & I\\I & 0 \end{array}\right)$,
$\bA=\cA_{r,r}^k$ and  

$\cA_{s,p}^k = L^k_sD^k_{s,p}U^k_p$ 
is a decomposition \eqref{eq_2} for matrix $\cA_{s,p}^k$. Then this decomposition has following properties:
\begin{equation}
\label{eq_5}
\hbox{if } \bC=0, \hbox{ then  }L^k_s=\diag(L^k_r,L^r_s).
\end{equation}

\begin{equation}
\label{eq_6}
\hbox{if } \bB=0, \hbox{ then  }U^k_p=\diag(U^k_r,U^r_p). 
\end{equation}

\begin{equation}
\label{eq_7}
\hbox{if } \bA=\bB=\bC=0, \bD=  L^r_s D^r_{s,p} U^r_s \hbox{ then  } \cA_{s,p}^k=S_2 
\left(\begin{array}{cc} L^r_s D^r_{s,p} U^r_s & 0\\0 & 0 \end{array}\right) S_2.
\end{equation}

\end{corollary} 
\begin{proof}
 Let $\bC=(\al^k_{i,j})_{i=r+1,..,s}^{j=k+1,..r}=0$.
 Then  all the minors $(\al^j_{i,j})$ for ${i=r+1,..,s}$ and ${j=k+1,..r}$ equal zero.
 It is follow from the  Sylvester Determinant Identity:
 $$
 \al^{j}_{i,j} * (\al^{k }_{k ,k })^{j-k-1}=\left|\begin{array}{cccc}
 \al^{k+1 }_{k+1,k+1} & ... &\al^{k+1 }_{k+1,j-1} & \al^{k+1 }_{k+1,j} \\
 ...  & ... & ...  &  ... \\
   \al^{k+1 }_{j-1,k+1} & ... &\al^{k+1 }_{j-1,j-1} & \al^{j-1 }_{k+1,j} \\
   \al^{k+1 }_{i,k+1} & ... &\al^{k+1 }_{i,j-1} & \al^{k+1 }_{i,j}
 \end{array}\right|,   \ k+1<j\leq i.  $$   
 The last row in this minor is a row in the block $\bC$.
 Therefore this row is zero row. So all the elements in the lower left corner block of the matrix $L^k_s$ equal zero.
 
 If $\bB=0$, then in the same manner we can prove that the upper right corner block of the matrix $U^k_p$ equals zero.
 
 If $\bA=\bB=\bC=0$, then   it is obvious that  the  last expression \eqref{eq_7} for $\cA_{s,p}^k$ is true.
\end{proof}

We use some block matrix notations:
for any matrix $A$ (or $A^p_q$) we denote by $A^{i_1,i_2}_{j_1,j_2}$
(or $A^{p;i_1,i_2}_{q;j_1,j_2}$) the block which stands at the intersection of rows  
$i_1+1, \ldots ,i_2 $ and columns  $j_1+1, \ldots ,j_2$  of the matrix.  
We denote by $A^{i_1 }_{i_2 }$ the diagonal block  $A^{i_1,i_2}_{i_1,i_2}$.

\subsection{ LDU Algorithm for the matrix with nonzero diagonal minors up to the rank}

 \noindent
{\it Input:} ($\cA_n^{k},\al^k$), $0\leq k<n$.

 \noindent
{\it Output:} $\{L^k_n$, $\{  \al^{k+1}, \al^{k+2}, \ldots , \al^n \}$, $U^k_n$, $M^k_n$, $W^k_n\}$, \ where
 $$D^k_n=\al^k\diag\{\al^k \al^{k+1}, \ldots , \al^{n-1} \al^{n}  \}^{-1}, M^k_n=\al^k(L^k_n D^k_n)^{-1}, W^k_n=\al^k(D^k_n U^k_n)^{-1}.$$
\noindent
1. If  $k=n-1$,  $\cA_n^{n-1}=(a^n)$ is a matrix of the first order, then we obtain
 $$\{ a^n, \{a^n\}, a^n, a^{n-1}, a^{n-1}  \}, \ \ D^{n-1}_n=(\al^n)^{-1}.$$

\noindent
2. If  $k=n-2$,  $\cA_n^{n-2}=\left(\begin{array}{cc} \al^{n-1} &\be\\\ga&\de \end{array}\right) $ is a matrix of second order, then we obtain

\noindent
$$
\bigg\{
\left(\begin{array}{cc} \al^{n-1}&0\\ \ga& \al^n
\end{array}\right),
 \{ \al^{n-1}, \al^{n} \},
\left(\begin{array}{cc}  \al^{n-1}&\be\\ 0&\al^n
\end{array}\right),
\left(\begin{array}{cc}\al^{n-2}&0\\- \ga& \al^{n-1}
\end{array}\right),
\left(\begin{array}{cc} \al^{n-2}&-\be\\ 0&\al^{n-1}
\end{array}\right)
\bigg\}
$$

\noindent
where $ \al^n= { (\al^{n-2})}^{-1} \Dmtr { \al^{n-1}} \be  \ga \de $,
  $D^{n-2}_n=\al^{n-2}\diag\{\al^{n-2} \al^{n-1},\al^{n-1} \al^{n}  \}^{-1}$.

\noindent 3. If the order of the matrix $ \cA^{k}_n  $ is more
than two ($0\leq k <n-2 $), then we choose an integer $ s $ in the
interval $ ( k <s <n) $ and divide the matrix into blocks
$$
\cA_n^{k}=
\left(\begin{array}{cc} \cA_s^{k} &\bB\\ \bC& \bD
\end{array}\right).
$$

\noindent
3.1. Recursive step: \ 
%
$
\{L^k_s, \{  \al^{k+1} ,  \al^{k+2}, \ldots , \al^s \},  U^k_s,
 M^k_s,\ W^k_s  \}
= \mathbf{LDU}(\cA_s^{k},\al^k)
$

\noindent
3.2.  We compute \ 
$
\widetilde U= (\al^k)^{-1} M^k_s\bB, \ \
   \widetilde L=  (\al^k)^{-1} \bC  W^k_s,
$
$$
\cA_n ^{s}=(\al^k)^{-1}\al^s( \bD -  \widetilde L D^k_s \widetilde U ).
$$
\noindent
3.3.   Recursive step:
$
\{L_n ^{s}, \{  \al^{s+1} ,  \al^{s+2}, \ldots , \al^n \},  U_n ^{s},  M^s_n,\ W^s_n  \}
= \mathbf{LDU}(\cA_n ^{s},\al^s )
$

\noindent
3.4 {\it Result:} \ 
$
\{L^k_n, \{  \al^{k+1} ,  \al^{k+2}, \ldots , \al^n \}, U^k_n, M^k_n, W^k_n  \},
$ 
where
$$
L^k_n=\left(\begin{array}{cc}   L^k_s&0\\ \widetilde L& L^s_n
\end{array}\right),
\
M^k_n =\left(\begin{array}{cc}   M^k_s  &0\\
 -   M^s_n \widetilde LD^k_s M^k_s /\al^{k} & \ \  M^s_n \ \
\end{array}\right),
$$
$$
 U^k_n= \left(\begin{array}{cc}   U^k_s& \widetilde U \\ 0& U^s_n
\end{array}\right), \
W^k_n = \left(\begin{array}{cc} \ \  W^k_s \ \  &  -  W^k_s D^k_s \widetilde U W^s_n/\al^{k} \\
 0 &   W^s_n
\end{array}\right).
$$

 Proof of the correctness of this algorithm   you can fined in \cite{18}.

 \section{Triangular matrix decomposition}
 
Now we are interesting the general case of initial matrix $A$ and we want to fined decomposition in the form 
\begin{equation}
\label{eq_8}
A^k_{n,m}=P^k_nL^k_nD^k_{n,m}U^k_mQ^k_m,
\end{equation}
with the additional 

\begin{property}[of triangular decomposition] 

If  matrix $A^k_{n,m}$ has rank $ {\bar r}-k $, then 
 

\noindent
($\alpha$) the lower and upper triangular matrices $L^k_n$ and $U^k_m$ are of the form
\begin{equation}
\label{eq_9}  
L^k_n= \left(\begin{array}{cc} L_1  &0\\ L_2 & I_{n-{\bar r}} \end{array}\right), \ 
U^k_m= \left(\begin{array}{cc} U_1  &U_2\\0&I_{m-{\bar r}} \end{array}\right),
\end{equation}
($\beta$)  the matrices $\cL= P^k_nL^k_n(P^k_n)^T $ and $\cU= (Q^k_m)^TU^k_mQ^k_m $ 
remain triangular after replacing in the  $L^k_n$ and 
$U^k_m$ of unit block  by arbitrary triangular block.
\end{property}
\begin{example}[ Non trivial example  for property $\beta$]
We replace of right lower unit block $I_2$ by the block $L_3$:
$ \left(\begin{array}{cc} 0 &I_1\\ I_2 & 0 \end{array}\right)
\left(\begin{array}{cc} L_1  &0\\ L_2 & \bL_3 \end{array}\right)
\left(\begin{array}{cc} 0 &I_2\\ I_1 & 0\end{array}\right), \hbox{ with } L_2=0.$
\end{example} 
\begin{example}[ Non trivial example  for property $\beta$]

\noindent
$ \left(\begin{array}{cc} P &0\\ 0 & I \end{array}\right)
\left(\begin{array}{cc} L_1  &0\\ L_2 & \bL_3 \end{array}\right)
\left(\begin{array}{cc} P^T &0\\ 0 & I\end{array}\right), \hbox{ with } PL_1P^T \hbox{lower triangular matrix.}$
\end{example} 
\begin{definition}[of triangular decomposition of matrix in domain]
The matrix decomposition \eqref{eq_8}
we call {\bf triangular decomposition} if it satisfies Property 1.
\end{definition} 
In this case, the decomposition \eqref{eq_8}  takes the form $A^k_{n,m}=\cL\cD\cU$ ($\cD=P^k_nD^k_{n,m}Q^k_n$).
 
This decomposition gives the Bruhat decomposition for the matrix $SA$ like it was shown in (4).
 
Let us note that the property 1 is trivial if $k=n-1$.
Property ($\beta$) is trivial if ${\bar r}\geq n-1$.
Therefore, the recursive algorithm as set out below, we can prove by induction.

\section{ LDU Algorithm with permutation matrices}

\subsection{Left upper block $\bA$ is not zero block}

 \noindent
{\it Input:} ($\cA_{n,m}^{k},\al^k$), $0\leq k<n$,  $0\leq k<n$, $\cA_{n,m}^{k}\neq 0$;

 \noindent
{\it Output:} $(P^k_n, L^k_n, \{  \al^{k+1} ,  \al^{k+2}, \ldots , \al^{\bar r} \},  U^k_m, Q^k_m, M^k_{\bar r}, W^k_{\bar r} )$,
  where  ${\bar r}-k$ is the rank of the  matrix $\cA_{n,m}^{k}$,  
$D^k_{\bar r}=\al^k\diag(\al^k \al^{k+1}, \ldots , \al^{{\bar r}-1} \al^{\bar r}  )^{-1}$,  
$D^k_{n,m}=\diag(D^k_{\bar r},0)$,  
$ M^k_{\bar r}=\al^k(L^k_{\bar r} D^k_{\bar r})^{-1}$,\ 
$W^k_{\bar r}=\al^k(D^k_{\bar r} U^k_{\bar r})^{-1}$, \ $A^k_n=P^k_nL^k_nD^k_{n,m}U^k_mQ^k_m$.
\bigskip

\noindent
1. If  $k=\min(n,m)-1$, $\cA_{n,m}^{k}$ is a matrix of one row or one column, then 
$$
\{I_n, L^k_n, \{  \al^{k+1} \},  U^k_m, I_m, (\al^{k+1}),\ (\al^{k+1})  \}
= \mathbf{LDU}(\cA_{n,m}^{k},\al^k).
$$
If $\cA_{n,m}^{k}=(a^{k+1},.., a^{k+1}_{n,{k+1}})^T$ is a column-matrix then $m=k+1$, $U^k_m=(a^{k+1})$ and 
$L^k_n=\left(\begin{array}{cc} & 0\\ \cA_{n,k+1}^{k} & I_{n-k}\end{array}\right)$.
If $\cA_{n,m}^{k}=(a^{k+1},.., a^{k+1}_{{k+1},m})$ is a row-matrix then $n=k+1$, $U^k_m=\left(\begin{array}{cc} &\cA_{k+1,m}^{k}\\
0  & I_{m-k}\end{array}\right)$ and $L^k_n=(a^{k+1})$. 
 
\bigskip
 
\noindent 2. If the  the matrix $\cA_{n,m}^{k} $ has more then one rows and  more then one columns, 
then we choose an integer $ s $ in the
interval $ ( k <s <n) $ and divide the matrix into blocks

\begin{equation}
\label{eq_10}
\cA_{n,m}^{k}=
\left(\begin{array}{cc} \cA_s^{k} &\bB\\ \bC& \bD
\end{array}\right).
\end{equation}
\noindent
2.1. Recursive step

Let the block $\cA_s^{k}$ has rank $r-k$, $r<s$, then we get  

$$
\{P^k_s, L^k_s, \{  \al^{k+1} ,  \al^{k+2}, \ldots , \al^{r} \},  U^k_s, Q^k_s,
 M^k_{r},\ W^k_{r}  \}
= \mathbf{LDU}(\cA_s^{k},\al^k)
$$
$L^k_s=\left(\begin{array}{cc} L^k_r  & 0\\ L_1 & I_{s-r}  \end{array}\right)$,\
$U^k_s=\left(\begin{array}{cc} U^k_r  &U_1\\ 0 &  I_{s-r} \end{array}\right)$,\
$D^k_s =\left(\begin{array}{cc} D^k_r  &0\\ 0 & 0  \end{array}\right)$,\
\
$M^k_{r}=\al^k    (L^k_r D^k_r)^{-1} $, 
$W^k_{r}= \al^k    (D^k_r U^k_r)^{-1}$, 
 $ D^k_r =\al^k\diag\{\al^k \al^{k+1}, \ldots , \al^{r-1} \al^{r}  \}^{-1}$,
 \ $A^k_s=P^k_sL^k_sD^k_sU^k_sQ^k_s$.

 Let us note that   $ \al^k_{i,j}$ and $\al^k$ is the minors of  the matrix $$(\diag(I_k, P^k_s)^T A (\diag(I_k,Q^k_s)^T).$$
 
\noindent
2.2. We separate the matrix $\cA_{n,m}^{k}$ into blocks another way 

\begin{equation}
\label{eq_11}
\diag^T ( P^k_s, I_{n-s})\cA_{n,m}^{k} \diag^T ( Q^k_s, I_{n-s})  =
\left(\begin{array}{cc} \cA_{r}^{k} & B\\  C&  D
\end{array}\right).
\end{equation}

\begin{equation}
\label{eq_12}
\widetilde U= (\al^k)^{-1} M^k_{r}B,
\widetilde L=  (\al^k)^{-1}  C  W^k_r,
\end{equation}

\begin{equation}
\label{eq_13}
\cA_{n,m}^r=(\al^k)^{-1}\al^r(  D -  \widetilde L D^k_{r} \widetilde U ).
\end{equation}

\noindent
2.3.   Recursive step.

If $A^r_{n,m} \neq 0$, then
$$
\{P_n^r,  L_n^{r},  \{  \al^{r+1} ,  \al^{r+2}, \ldots , \al^{\bar r} \},  U_m ^{r}, Q_m^r, M^r_{\bar r},\ W^r_{\bar r}  \}
= \mathbf{LDU}(\cA_{n,m} ^{r},\al^r )
$$
$$A^r_{n,m}=P^r_nL^r_nD^r_{n,m}U^r_mQ^r_m.$$

Otherwise($A^r_{n,m} = 0$), we do nothing.
\\
2.4 {\it Result:}
$$
\{P^k_n,L^k_n, \{  \al^{k+1} ,  \al^{k+2}, \ldots , \al^{\bar r} \}, U^k_m, Q^k_m, M^k_{\bar r}, W^k_{\bar r}  \},
$$

If  $A^r_{n,m} \neq 0$, then $$P^k_n=\diag(P^k_s, I_{n-s})\diag(I_{r-k}, P^r_n),Q^k_m=\diag(I_{r-k}, Q^r_m)\diag(Q^k_s, I_{n-s})$$,
\begin{equation}
\label{eq_14}
L^k_n=\left(\begin{array}{cc}   L^k_r&0\\ {P^r_n}^T \widetilde L& L^r_n
\end{array}\right),
\
 U^k_m= \left(\begin{array}{cc}   U^k_r& \widetilde U {Q^r_n}^T\\ 0& U^r_n
\end{array}\right),
\end{equation}

\begin{equation}
\label{eq_15}
M^k_{\bar r} =\left(\begin{array}{cc}   M^k_r  &0\\
 -   M^r_{\bar r} {P^r_n}^T \widetilde LD^k_r M^k_r /\al^{k} & \ \  M^r_{\bar r} \ \
\end{array}\right),
\end{equation}
\begin{equation}
\label{eq_16}
 W^k_{\bar r} = \left(\begin{array}{cc} \ \  W^k_r \ \  &  -  W^k_r D^k_r \widetilde U {Q^r_m}^T W^r_{\bar r}/\al^{k} \\
 0 &   W^r_{\bar r}
\end{array}\right).
\end{equation}

Otherwise($A^r_{n,m} = 0$), 
$$\bar{r} = r$$
$$P^k_n=\diag(P^k_s, I_{n-s}), Q^k_m=\diag(Q^k_s, I_{n-s}),$$
\begin{equation}
\label{eq_17}
L^k_n=\left(\begin{array}{cc}   L^k_r&0\\ \widetilde L& I_{n-r}
\end{array}\right),
\
 U^k_m= \left(\begin{array}{cc}   U^k_r& \widetilde U \\ 0& I_{m-r}
\end{array}\right),
\end{equation}
$$M^k_{r}=\al^k    (L^k_r D^k_r)^{-1},$$
$$W^k_{r}= \al^k    (D^k_r U^k_r)^{-1},$$ 
$$D^k_r =\al^k\diag\{\al^k \al^{k+1}, \ldots , \al^{r-1} \al^{r}  \}^{-1},$$

\begin{proof}

Proof of this theorem is like previous theorem. But now we have to proof that matrices $P^k_n, Q^k_m, L^k_n, U^k_m$
 satisfy the Property of triangular decomposition.
 
 Let matrices $P^k_s, Q^k_s, L^k_s, U^k_s$ and $P^r_n, Q^r_n, L^r_n, U^r_n$
 satisfy the  Property of triangular decomposition and $\cA^r_{n,m} \neq 0$ in the step 2.3.
 
 {\it   Property $\beta$}.
 We wish to proof property $(\beta)$ for $P^k_n  L^k_n (P^k_n)^T$ and $(Q^k_m)^T  U^k_m  Q^k_m $.
\begin{align} 
P^k_n  L^k_n (P^k_n)^T & = \diag(P^k_s, I_{n-s})\diag(I_{r-k}, P^r_n)  L^k_n \diag^T(I_{r-k}, P^r_n) \diag^T(P^k_s, I_{n-s}) =
\nonumber\\
\label{eq_18} 
&= \diag(P^k_s, I_{n-s}) \left(\begin{array}{cc}   L^k_r&0\\ \widetilde L&  {P^r_n}   L^r_n {P^r_n}^T 
\end{array}\right)\diag(P^k_s, I_{n-s})^T.
\end{align}
The blocks ${P^r_n} L^r_n {P^r_n}^T$ and 
\begin{equation}
\label{eq_19}
P^k_s L^k_s (P^k_s)^T=
  P^k_s\left(\begin{array}{cc} L^k_r  & 0\\ L_1 & I  \end{array}\right) (P^k_s)^T
\end{equation}
are triangular by induction.
The block 

\begin{equation}
\label{eq_20}
G= P^k_s  \left(\begin{array}{cc}   L^k_r &0 \\ {\widetilde L}'  &  L''
\end{array}\right) (P^k_s )^T
\end{equation}
is located in the upper left corner of the matrix \eqref{eq_18}.
The block ${\widetilde L}'$ has size $ (s-r)\times(r-k)$ and $L''$  has size $ (s-r)\times(s-r)$. 
The block $L''$ is low triangular block and ${\widetilde L}'= L_1$ by the construction.
Due to the  property ($\beta$) of triangular decomposition of $A^k_s$ this corner block $G$ is triangular too.

We can make a similar argument for the block $(Q^k_m)^T  U^k_m  Q^k_m$.

{\it  Property $\alpha$}.  Due to the construction of the matrices $L^k_n$ and $U^k_m$ they can have the unit block in the lower right corner,
but the size of this block is less then $n-r$. Therefore this unit block is disposed in the blocks $L^r_n$ and $U^r_n$. 
The property ($\alpha$) of triangular decomposition of $A^k_n$ is the consequence of such property of matrix $A^{r}_n$.
The case when $\cA^r_{n,m}=0$ in the step 2.3 is evident.
\end{proof}

{\it Collary.}

In section 6 we present a full example of triangular decomposition for a matrix of size 6$\times$6.
In this example the \eqref{eq_18}-\eqref{eq_20} can be written as follows:
\begin{align}
\eqref{eq_18}:\ 
& P^0_6 L^0_6 {P^0_6}^T = \diag(P^0_4, I_{2})\diag(I_{2}, P^2_6)  L^0_6 \diag^T(I_{2}, P^2_6) \diag^T(P^0_4, I_{2}) = 
\nonumber\\
& = \diag(P^0_4, I_{2})
\left(\begin{array}{cc}
L^0_2&0\\
\widetilde L^0_2 & P^2_6 L^2_6 {P^2_6}^T 
\end{array}\right)
\diag(P^0_4, I_{2})^T,
\nonumber\\
\eqref{eq_19}:\ 
&P^0_4 L^0_4 (P^0_4)^T =
P^0_4\left(\begin{array}{cc} L^0_2  & 0\\ \overline{L^0_2} & I_2  \end{array}\right) (P^0_4)^T,
\nonumber\\
\eqref{eq_20}:\ 
& G=
\left(\begin{array}{cc}
L^0_2 & 0 \\
\overline L^0_2 & L^4_6
\end{array}\right)
\nonumber.
\end{align}

\subsection{ Matrix A has full rank, matrices $C$  and (or) $B$ are zero matrices.}

\noindent
2.1. This step has no changes.
 
\noindent
2.2.  We compute $\widetilde U$,  $\widetilde L$. One of these matrices or both equal zero.
So for matrix $\cA_n ^{s}$ we obtain: $\cA_n ^{s}=(\al^k)^{-1}\al^s\bD $.

\noindent
2.3.   Recursive step. 
We take matrix 	$\bD$ instead of $\cA_n ^{s}$. 
$$
\{P_n ^{s}, \bar L_n ^{s}, \{  \bar \al^{s+1} ,  \bar \al^{s+2}, \ldots ,\bar \al^n \}, \bar U_n ^{s}, Q_n ^{s}, \bar M^s_n,\ \bar W^s_n  \}
= \mathbf{LDU}(\bD,\al^k )
$$
Let us note that we can do simultaneously this computations and the computations of the first step (3.1).

The value $\lambda=\al^s/\al^k$ is a coefficient in the equality  $\cA_n ^{s}=(\al^k)^{-1}\al^s\bD$. So it is ease to proof that 
$\lambda \bar L_n^{s}= L_n ^{s}$, 
$\lambda\bar \al^{s+1}= \al^{s+1},  \lambda \bar \al^{s+2}= \al^{s+2}, \ldots ,
\lambda \bar \al^n = \al^n$, $\lambda^{-1} \bar D_n^{s}=  D_n^{s}$,
$ \lambda\bar U_n ^{s}=  U_n^{s},\lambda \bar M^s_n=  M^s_n, \lambda \bar W^s_n=W^s_n$.
This way we can fined factorization:
$$
\{P_n ^{s}, L_n ^{s}, \{  \al^{s+1} ,  \al^{s+2}, \ldots , \al^n \},  U_n ^{s}, Q_n ^{s},  M^s_n,\ W^s_n  \}
= \mathbf{LDU}(\cA_n ^{s},\al^s )
$$
\noindent
2.4 Result has no changes. But if $C=0$ then  
$L^k_n=\diag(  L^k_s,  L^s_n),$ 
  if $B=0$ then  $U^k_m=\diag(   U^k_s, U^s_n)$.

\subsection{ Matrix A has no full rank, matrices $C$  and (or) $B$ are zero matrices, $A \neq 0$.}


\noindent  Let us at some stage of algorithm 4.1 we obtain the matrix 

\begin{equation}
\cA^k_{n,m}=
\left(\begin{array}{cc} \bA  &\bB\\ 0& \bD
\end{array}\right).
\nonumber
\end{equation}
Here $\bA= \cA^{k}_s  $ and  $\bD$ is a block of size $(n-s)\times(m-s)$. 

 We can use the algorithm, which is set out in subsection 4.1. 
 But in this case, the decomposition of block $\bA=\cA_s^{k}$ and $\bD$ can be performed simultaneously and independently.

\noindent
2.1. Recursive steps
%
\begin{equation}
\label{eq_21}
\{P^k_s, L^k_s, \{  \al^{k+1} ,  \al^{k+2}, \ldots , \al^r \},  U^k_s, Q^k_s
 M^k_r,\ W^k_r ,  \}
= \mathbf{LDU}(\bA,\al^k)
\end{equation}
\begin{equation}
\label{eq_22}
\{P_n^{s}, \bar L_n ^{s}, \{  \bar \al^{s+1} ,  \bar \al^{s+2}, \ldots ,\bar \al^{\bar r} \}, 
\bar U_m ^{s}, Q_m ^{s}, \bar M^s_{\bar r},\ \bar W^s_{\bar r}  \}
= \mathbf{LDU}(\bD,\al^k )
\end{equation}
$\bA = P_s ^{k}L_s ^{k} D_s ^{k} U_s ^{k} Q_s ^{k}, \ \bD = P_m ^{s} \bar L_n ^{s} D_{n,m}^{s} \bar U_m ^{s} Q_m ^{s}.$
 
We suppose that matrix $\bA $ have rank $r-k$, $k<r<s$, 
matrix $\bD $ have rank ${\bar r}-s$, $s<{\bar r}\leq\min(n,m)$. 
Then  
  $$
  L_s ^{k}=\left(\begin{array}{cc} L_0  & 0\\ M_0 & I_{s-r}  \end{array}\right) ,
 D_s ^{k}=\left(\begin{array}{cc} d_1  &0\\ 0 & 0  \end{array}\right), 
  U_s ^{k}=\left(\begin{array}{cc} U_0  &V_0\\ 0 & I_{s-r}   \end{array}\right),
 $$
\noindent
2.2. We separate the matrix $\cA_{n,m}^{k}$ into blocks another way 

\begin{equation}
\nonumber
\diag^T ( P^k_s, I_{n-s})\cA_{n,m}^{k} \diag^T ( Q^k_s, I_{n-s})  =
\left(\begin{array}{cc} \cA_{r}^{k} & B\\  C&  D
\end{array}\right).
\end{equation}

\begin{equation}
\label{eq_23}
\widetilde U= (\al^k)^{-1} M^k_{r}B, 
\widetilde L=  (\al^k)^{-1}  C  W^k_r,
\end{equation}

\begin{equation}
\label{eq_24}
\cA_{n,m}^r=(\al^k)^{-1}\al^r(  D -  \widetilde L D^k_{r} \widetilde U ).
\end{equation}

\noindent
Let us note that $\lambda=\al^{r}/\al^{k}$,  
$C=\left(\begin{array}{c} C_1  \\ 0 \end{array}\right)$,
$\widetilde L= \left(\begin{array}{c} L_1  \\ 0 \end{array}\right)$,
$\cA_{n,m}^r=\lambda \left(\begin{array}{cc} D_1&D_2  \\ 0&\bD \end{array}\right)$.

\noindent
2.3.   Recursive step.
 
$$
\{P_n^r,  L_n^{r},  \{  \al^{r+1} ,  \al^{r+2}, \ldots , \al^{\bar r} \},  U_m ^{r}, Q_m^r, M^r_{\bar r},\ W^r_{\bar r}  \}
= \mathbf{LDU}(\cA_{n,m} ^{r},\al^r )
$$
we can do with the help of block decomposition \eqref{eq_22}, and obtain

\ $A^r_{n,m}=P^r_nL^r_nD^r_{n,m}U^r_mQ^r_m$.

\noindent
2.4 {\it Result:}
$$
\{P^k_n,L^k_n, \{  \al^{k+1} ,  \ldots , \al^{r}, \al^{r+1}, \ldots , \al^{\tilde r} \}, U^k_m, Q^k_m, M^k_{\tilde r}, W^k_{\tilde r}  \}
=\mathbf{LDU}(\cA_{n,m}^{k},\al^k),
$$ 
is similar as in the section 4.1.

\section{ Matrix A is zero matrix }

\subsection{ Matrix A is zero matrix,   $C$  and (or) $B$ are nonzero matrices }
\noindent  Let us at some stage of algorithm 4.1 we obtain the matrix 

\begin{equation}
\label{eq_25}
 \cA^k_{n,m}=
\left(\begin{array}{cc} 0  &\bB\\ \bC& \bD
\end{array}\right), \ S=\left(\begin{array}{cc} 0  & I \\ I & 0 \end{array}\right).
\end{equation}
Let $\bC\neq 0$.
Then for the matrix $S \cA^k_{n,m}$ we can make triangular decomposition as in section 4:
$\left( \begin{array}{cc}\bC & \bD \\ 0 & \bB \end{array}\right)  = LDU $. Then $ \cA^k_{n,m}=S LDU $. 
In this case martix L has form $L=\diag(L_1,L_2)$ due to corollary 3 of theorem 1. Then we get $\cA^k_{n,m}=\diag(L_2,L_1)(SD)U$. 

Let $\bB\neq 0$.
Then for the matrix $\cA^k_{n,m} S$ we can make triangular decomposition as in section 4:
$\left( \begin{array}{cc}\bB & 0 \\ \bD & \bC \end{array}\right)  = LDU $. Then $ \cA^k_{n,m}=LDUS $. 
In this case martix U has form $U=\diag(U_1,U_2)$ due to corollary 3 of theorem 1. Then we get $\cA^k_{n,m}=L(DS)\diag(U_2,U_1)$. 

Let $\bB=\bC=\bA=0$ and $\bD\neq 0$, $\bD=LDU$.
Then for the matrix $ \cA^k_{n,m} $ we can make triangular decomposition  :
$\left( \begin{array}{cc}0 & 0 \\ 0 & \bD \end{array}\right)  = 
\left( \begin{array}{cc}I & 0 \\ 0 & L \end{array}\right) \left( \begin{array}{cc}0 & 0 \\ 0 & D \end{array}\right)
\left( \begin{array}{cc}I & 0 \\ 0 & U \end{array}\right) 
$.

\subsection{The all cases when half of the matrix  is equal zero}

\begin{theorem}
Let $\cA_{n,m}^{k}=
\left(\begin{array}{cc} \bA &\bB\\ \bC& \bD
\end{array}\right).$

If $\bC=\bD=0$ and  $(\bA,\bB)=PL(D,0)UQ$ is a triangular decomposition   then 
$ 
\left(\begin{array}{cc } P&0 \\ 0 &  I \end{array}\right)
\left(\begin{array}{cc } L&0 \\ 0 &  I \end{array}\right)
\left(\begin{array}{cc } D&0 \\ 0 &  0 \end{array}\right)
UQ $
is a triangular decomposition of $\cA_{n,m}^{k}$.

If $\bA=\bB=0$ and $(\bC,\bD)=PL(D,0)UQ$ is a triangular decomposition   then 
$
\left(\begin{array}{cc } 0&I \\ P &  0 \end{array}\right)
\left(\begin{array}{cc } L&0 \\ 0 &  I \end{array}\right)
\left(\begin{array}{cc } D&0 \\ 0 &  0 \end{array}\right)
UQ $
is a triangular decomposition of $\cA_{n,m}^{k}$.
 
If $\bB=\bD=0$ and $(\bA,\bC)^T=PL(D,0)^TUQ$ is a triangular decomposition then
$ 
PL
\left(\begin{array}{cc } D&0 \\ 0 &  0 \end{array}\right)
\left(\begin{array}{cc } U&0 \\ 0 &  I \end{array}\right)
\left(\begin{array}{cc } Q&0 \\ 0 &  I \end{array}\right)
$
is a triangular decomposition of $\cA_{n,m}^{k}$.

If $\bA=\bC=0$ and $(\bB,\bD)^T=PL(D,0)^TUQ$ is a triangular decomposition   then 
$ 
PL
\left(\begin{array}{cc } D&0 \\ 0 &  0 \end{array}\right)
\left(\begin{array}{cc } U&0 \\ 0 &  I \end{array}\right)
\left(\begin{array}{cc } 0&I \\ Q &  0 \end{array}\right)
$
is a triangular decomposition of $\cA_{n,m}^{k}$.

\end{theorem}
\begin{proof} To proof we have to check the Property of triangular decomposition for each cases.
\end{proof}

\section{Example of triangular decomposition on $\mathbb{Z}$.}
Given a matrix $A$. We show how to obtain the decomposition  $A=\bL\bD\bU$:
$$
\left[
\begin{array}{cccccc}
3&2&3&5&1&2\\
1&3&4&2&3&4\\
3&2&3&5&5&6\\
1&3&4&2&2&1\\
2&1&3&2&2&3\\
2&1&3&2&2&3\\
\end{array}
\right]
=
\left[
\begin{array}{cccccc}
3& 0&  0&   0&   0&  0\\
1& 7&  0&   0&   0&  0\\
3& 0&  40&  0&   0&  0\\
1& 7&  -10& -80& 0&  0\\
2& -1& 0&   0&   10& 0\\
2& -1& 0&   0&   10& 1
\end{array}
\right]
\left[
\begin{array}{cccccc}
\frac{1}{3}& 0&      0&      0& 0&       0        \\
0&     \frac{1}{21}& 0&      0& 0&       0        \\
0&     0&      0&      0& \frac{1}{400}& 0        \\
0&     0&      0&      0& 0&       \frac{-1}{3200}\\
0&     0&      \frac{1}{70}& 0& 0&       0        \\
0&     0&      0&      0& 0&       0        \\
\end{array}
\right]
\left[
\begin{array}{cccccc}
3& 2& 3&  5&  1&  2  \\
0& 7& 9&  1&  8&  10 \\
0& 0& 10& -9& 12& 15 \\
0& 0& 0&  1&  0&  0  \\
0& 0& 0&  0&  40& 40 \\
0& 0& 0&  0&  0&  -80\\ 
\end{array}
\right]
$$
Let the upper left block has a size of $4 \times 4$:
 
\noindent
$\cA^0_6 = 
\left(
\begin{array}{cc}
\cA^0_4&\cB^0_4\\
\cC^0_4&\cD^0_4
\end{array}
\right)
$,
$\cA^0_4 = 
\left[
\begin{array}{cccc}
3&2&3&5\\
1&3&4&2\\
3&2&3&5\\
1&3&4&2\\
\end{array}
\right]$,
$\cB^0_4 = \left(
\begin{array}{cc}
1&2\\
3&4\\
\end{array}
\right)$,
$\cC^0_4=\left(
\begin{array}{cccc}
2&1&3&2\\
2&1&3&2\\
\end{array}
\right)$,
$\cD^0_4 = \left(
\begin{array}{cc}
2&3\\
2&3\\
\end{array}
\right)$.
\subsection{Recursive step of first level. Triangular decomposition for $\cA^0_4$.}
Let us split matrix $\cA^0_4$ into blocks again.
$\cA^0_4 = 
\left(
\begin{array}{cc}
\cA^0_2&\cB^0_2\\
\cC^0_2&\cD^0_2
\end{array}
\right)
$,
where $\cA^0_2 = 
\left(
\begin{array}{cc}
3&2\\
1&3\\
\end{array}
\right)$,
$\cB^0_2 = \left(
\begin{array}{cc}
3&5\\
4&2\\
\end{array}
\right)$,
$\cC^0_2=\left(
\begin{array}{cc}
3&2\\
1&3\\
\end{array}
\right)$,
$\cD^0_2 = \left(
\begin{array}{cc}
3&5\\
4&2\\
\end{array}
\right)$.
\subsubsection{6.1.1 Recursive step. Triangular decomposition for $\cA^0_2$.}
We split matrix $\cA^0_2$:
$\cA^0_2 = 
\left(
\begin{array}{cc}
\cA^0_1&\cB^0_1\\
\cC^0_1&\cD^0_1
\end{array}
\right)
$,
where $\cA^0_1 = 
\left(
\begin{array}{c}
3
\end{array}
\right)$,
$\cB^0_1 = \left(
\begin{array}{c}
2\\
\end{array}
\right)$,
$\cC^0_1=\left(
\begin{array}{c}
1
\end{array}
\right)$,
$\cD^0_1 = \left(
\begin{array}{c}
3
\end{array}
\right)$.
\subsubsection{6.1.1.1 Recursive step. Triangular decomposition for $\cA^0_1$.}
Matrix $\cA^0_1$ has size $1 \times 1$, then we can write result for this decomposition:
$\cA^0_1 = 
\left(
\begin{array}{c}
3
\end{array}
\right)$,$\al_0 = 1$,
$$\left\lbrace P^0_1,L^0_1,\lbrace\al_1\rbrace,U^0_1,Q^0_1,M^0_1,W^0_1 \right\rbrace = \mathbf{LDU}(\cA_1^{0},\al_0),$$
$L^0_1 =  U^0_1 = (3)$, $P^0_1 = Q^0_1 =  M^0_1 = W^0_1 = (1)$, $\al_1 = 3$, $D^0_1 = (\frac{1}{3})$.
\subsubsection{Calculations for next recursive step.}
\begin{align}
\overline{L^0_1} & = (\al_0)^{-1}\cC^0_1 Q^0_1 W^0_1= (1)^{-1}(1)(1)(1) = (1),\nonumber\\
\overline{U^0_1} & = (\al_0)^{-1}M^0_1 P^0_1 \cB^0_1= (1)^{-1}(1)(1)(2) = (2),
\nonumber\\
\cA^1_2 & =  ({\al_1/\al_0})(\cD^0_1 - \overline{L^0_1} D^0_1 \overline{U^0_1}) = \frac{3}{1}((3)-(1)( {1/3})(2)) = (7).
\nonumber
\end{align}
\subsubsection{6.1.1.2 Recursive step. Triangular decomposition for $\cA^1_2$.}
Matrix $\cA^1_2$ has size $1 \times 1$: 
$\cA^1_2 = 
\left(
\begin{array}{c}
7
\end{array}
\right)$, $\al_1 = 3$,
$$\left\lbrace P^1_2,L^1_2,\lbrace\al_2\rbrace,U^1_2,Q^1_2,M^1_2,W^1_2 \right\rbrace = \mathbf{LDU}(\cA^1_2,\al^1),$$
$L^1_2 = U^1_2 = (7)$,
$P^1_2 = Q^1_2 = (1)$,
$M^1_2 =  W^1_2 = (3)$, $\al_2 = 7$,
$D^1_2 = ( {1/7})$.
$$\overline{M^0_1} = - (\al_0)^{-1} M^1_2 \overline{L^0_1} D^0_1 M^0_1 = (1)^{-1} (3)(1)( {1/3}) (1) = (-1) $$
$$\overline{W^0_1} = - (\al_0)^{-1} W^0_1 D^0_1 \overline{U^0_1} W^1_2 = (1)^{-1} (1)({1/3}) (2) (3) = (-2) $$
\subsubsection{6.1.1.3 Result of decomposition of matrix $\cA^0_2$.}
$$\left\lbrace P^0_2,L^0_2,\lbrace\al_1,\al_2\rbrace,U^0_2,Q^0_2,M^0_2,W^0_2 \right\rbrace = \mathbf{LDU}(\cA^0_2,\al_0),$$
$
P^0_2 = 
\left(
\begin{array}{cc}
P^0_1&0\\
0&P^1_2\\
\end{array}
\right)
=
Q^0_2 =
\left(
\begin{array}{cc}
Q^0_1&0\\
0&Q^1_2\\
\end{array}
\right)
=
\left(
\begin{array}{cc}
1&0\\
0&1\\
\end{array}
\right)
$,
$
L^0_2 = 
\left(
\begin{array}{cc}
L^0_1&0\\
\overline{L^0_1}&L^1_2\\
\end{array}
\right)
=
\left(
\begin{array}{cc}
3&0\\
1&7\\
\end{array}
\right)
$,
$
U^0_2 =
\left(
\begin{array}{cc}
U^0_1&\overline{U^0_1}\\
0&U^1_2\\
\end{array}
\right)
=
\left(
\begin{array}{cc}
3&2\\
0&7\\
\end{array}
\right)
$,
$
M^0_2 =
\left(
\begin{array}{cc}
M^0_1&0\\
\overline{M^0_1}&M^1_2\\
\end{array}
\right)
=
\left(
\begin{array}{cc}
1&0\\
-1&3\\
\end{array}
\right)
$,
$\al_1 = 3$, $\al_2=7$, 
$
W^0_2 =
\left(
\begin{array}{cc}
W^0_1&\overline{W^0_1}\\
0&W^1_2\\
\end{array}
\right)
=
\left(
\begin{array}{cc}
1&-2\\
0&3\\
\end{array}
\right)
$,
$
D^0_2 = 
\left(
\begin{array}{cc}
 {1/3}&0\\
0& {1/21}\\
\end{array}
\right)
$.
\subsubsection{Calculations for next recursive step.}
\begin{align}
\overline{L^0_2} & = (\al_0)^{-1}\cC^0_2 Q^0_2 W^0_2= (1)^{-1}
\left(
\begin{array}{cc}
3&2\\
1&3\\
\end{array}
\right)
\left(
\begin{array}{cc}
1&0\\
0&1\\
\end{array}
\right)
\left(
\begin{array}{cc}
1&-2\\
0&3\\
\end{array}
\right) = 
\left(
\begin{array}{cc}
3&0\\
1&7\\
\end{array}
\right),
\nonumber\\
\overline{U^0_2} & = 
(\al_0)^{-1}M^0_2 P^0_2 \cB^0_2= 
(1)^{-1}
\left(
\begin{array}{cc}
1&0\\
-1&3\\
\end{array}
\right)
\left(
\begin{array}{cc}
1&0\\
0&1\\
\end{array}
\right)
\left(
\begin{array}{cc}
3&5\\
4&2\\
\end{array}
\right) = 
\left(
\begin{array}{cc}
3&5\\
9&1\\
\end{array}
\right),
\nonumber\\
\cA^2_4 & = \frac{\al_2}{\al_0}(\cD^0_2 - \overline{L^0_2} D^0_2 \overline{U^0_2}) = \frac{7}{1}
\left(
\left(
\begin{array}{cc}
3&5\\
4&2\\
\end{array}
\right)
-
\left(
\begin{array}{cc}
3&0\\
1&7\\
\end{array}
\right)
\left(
\begin{array}{cc}
\frac{1}{3}&0\\
0&\frac{1}{21}\\
\end{array}
\right)
\left(
\begin{array}{cc}
3&5\\
9&1\\
\end{array}
\right)
\right) = 
\left(
\begin{array}{cc}
0&0\\
0&0\\
\end{array}
\right).
\nonumber
\end{align}
\subsubsection{6.1.2 Recursive step. Triangular decomposition for $\cA^2_4$} 
Since the matrix $\cA^2_4$ is zero, we can write the result of decomposition of matrix $\cA^0_4$.
\subsubsection{6.1.3 Result of decomposition of matrix $\cA^0_4$}
$$\left\lbrace P^0_4,L^0_4,\lbrace\al_1,\al_2\rbrace,U^0_4,Q^0_4,M^0_2,W^0_2 \right\rbrace = \mathbf{LDU}(\cA^0_4,\al^0),$$
$P^0_4 = 
\left(
\begin{array}{cc}
P^0_2&0\\
0&I_2\\
\end{array}
\right)
 $,
$L^0_4 =
\left(
\begin{array}{cc}
L^0_2&0\\
\overline{L^0_2}&I_2\\
\end{array}
\right) $,
$
U^0_4 =
\left(
\begin{array}{cc}
U^0_2&\overline{U^0_2}\\
0&I_2\\
\end{array}
\right)
 $,
$
Q^0_4 =
\left(
\begin{array}{cc}
Q^0_2&0\\
0&I_2\\
\end{array}
\right)
 $,
$
M^0_2 =
\left(
\begin{array}{cc}
1&0\\
-1&3\\
\end{array}
\right)
$,
$
W^0_2 =
\left(
\begin{array}{cc}
1&-2\\
0&3\\
\end{array}
\right)
$,
$
D^0_4 = 
\left(
\begin{array}{cc}
D^0_2&0\\
0&0\\
\end{array}
\right)
 $,
$\al_1 = 3$,
$\al_2=7$, 
$(P^0_4, L^0_4, D^0_4, U^0_4, Q^0_4)=$ 
$
\left(\left(
\begin{array}{cccc}
1&0&0&0\\
0&1&0&0\\
0&0&1&0\\
0&0&0&1\\
\end{array}
\right),
\left(
\begin{array}{cccc}
3&0&0&0\\
1&7&0&0\\
3&0&1&0\\
1&7&0&1\\
\end{array}
\right),
\left(
\begin{array}{cccc}
\frac{1}{3}&0&0&0\\
0&\frac{1}{21}&0&0\\
0&0&0&0\\
0&0&0&0\\
\end{array}
\right),
\left(
\begin{array}{cccc}
3&2&3&5\\
0&7&9&1\\
0&0&1&0\\
0&0&0&1\\
\end{array}
\right),
\left(
\begin{array}{cccc}
1&0&0&0\\
0&1&0&0\\
0&0&1&0\\
0&0&0&1\\
\end{array}
\right)\right).
$

\subsection{Recursive step of first level.}
Since   $\cA^0_4$ is not full rank, we should  split matrix   again:
$P^0_4 \cA^0_6 Q^0_4=
\left(
\begin{array}{cc}
\cA^0_2&\cB^0_2\\
\cC^0_2&\cD^0_2
\end{array}
\right),
$
where $\cA^0_2 = 
\left(
\begin{array}{cc}
3&2\\
1&3\\
\end{array}
\right)$,
$\cB^0_2 = \left(
\begin{array}{cccc}
3&5&1&2\\
4&2&3&4\\
\end{array}
\right)$,
$\cC^0_2=\left(
\begin{array}{cc}
3&2\\
1&3\\
2&1\\
2&1\\
\end{array}
\right)$
$\cD^0_2 = \left(
\begin{array}{cccc}
3&5&5&6\\
4&2&2&1\\
3&2&2&3\\
3&2&2&3\\
\end{array}
\right)$.
\\
$
\widetilde{L^0_2} = (\al_0)^{-1}\cC^0_2 W^0_2  = (1)^{-1}
\left(
\begin{array}{cc}
3&2\\
1&3\\
2&1\\
2&1\\
\end{array}
\right)
\left(
\begin{array}{cc}
1&-2\\
0&3\\
\end{array}
\right) = 
\left(
\begin{array}{cc}
3&0\\
1&7\\
2&-1\\
2&-1\\
\end{array}
\right), 
\widetilde{U^0_2}   =  
(\al_0)^{-1}M^0_2 \cB^0_2 =  
(1)^{-1}
\left(
\begin{array}{cc}
1&0\\
-1&3\\
\end{array}
\right)
\left(
\begin{array}{cccc}
3&5&1&2\\
4&2&3&4\\
\end{array}
\right)= 
\left(
\begin{array}{cccc}
3&5&1&2\\
9&1&8&10\\
\end{array}
\right),
\cA^2_6   =  \frac{\al_2}{\al_0}(\cD^0_2 - \widetilde{L^0_2} D^0_2 \widetilde{U^0_2}) =\\
\frac{7}{1}
\left(
\left(
\begin{array}{cccc}
3&5&5&6\\
4&2&2&1\\
3&2&2&3\\
3&2&2&3\\
\end{array}
\right)
-
\left(
\begin{array}{cc}
3&0\\
1&7\\
2&-1\\
2&-1\\
\end{array}
\right)
\left(
\begin{array}{cc}
\frac{1}{3}&0\\
0&\frac{1}{21}\\
\end{array}
\right)
\left(
\begin{array}{cccc}
3&5&1&2\\
9&1&8&10\\
\end{array}
\right)
\right) =
\left(
\begin{array}{cccc}
0&0&28&28\\
0&0&-7&-21\\
10&-9&12&15\\
10&-9&12&15\\
\end{array}
\right). 
$

\noindent
Split matrix $\cA^2_6$ into blocks.
$\cA^2_6 = 
\left(
\begin{array}{cc}
\cA^2_4&\cB^2_4\\
\cC^2_4&\cD^2_4
\end{array}
\right)$,
where $\cA^2_4 = 
\left(
\begin{array}{cc}
0&0\\
0&0\\
\end{array}
\right)$,
$\cB^2_4 = \left(
\begin{array}{cc}
28&28\\
-7&-21\\
\end{array}
\right)$,
$\cC^2_4=\left(
\begin{array}{cc}
10&-9\\
10&-9\\
\end{array}
\right)$,
$\cD^2_4 = \left(
\begin{array}{cc}
12&15\\
12&15\\
\end{array}
\right)$.
Since matrix $\cA^2_4$ is zero, we should use some permutation of blocks and
then we can find decomposition of $\cC^2_4$ and $\cB^2_4$ simultaneously.
$$
\left(
\begin{array}{cc}
0&I_2\\
I_2&0\\
\end{array}
\right)
\cA^2_6 = 
\left(
\begin{array}{cc}
0&I_2\\
I_2&0\\
\end{array}
\right)
\left(
\begin{array}{cc}
0&\cB^2_4\\
\cC^2_4&\cD^2_4
\end{array}
\right)=
\left(
\begin{array}{cc}
\cC^2_4&\cD^2_4\\
0&\cB^2_4\\
\end{array}
\right).$$ 
\subsubsection{6.2.1 Recursive step. Triangular decomposition of matrix.}
$$\cC^2_4=\left(
\begin{array}{cc}
10&-9\\
10&-9\\
\end{array}
\right),\al_2=7$$
$$\left\lbrace P^2_4,L^2_4,\lbrace\al_3\rbrace,U^2_4,Q^2_4,M^2_3,W^2_3 \right\rbrace = \mathbf{LDU}(\cC^2_4,\al^2),$$ where 
$P^2_4 = 
\left(
\begin{array}{cc}
1&0\\
0&1\\
\end{array}
\right)$,
$L^2_4 = \left(
\begin{array}{cc}
L_0&0\\
M_0&I_1\\
\end{array}
\right)=
\left(
\begin{array}{cc}
10&0\\
10&1\\
\end{array}
\right)$,
$U^2_4=\left(
\begin{array}{cc}
U_0& V_0\\
0&I_1\\
\end{array}
\right)=\left(
\begin{array}{cc}
10& -9\\
0& 1\\
\end{array}
\right)$,
$Q^2_4 = 
\left(
\begin{array}{cc}
1&0\\
0&1\\
\end{array}
\right)$,
$M^2_3 = 
\left(
\begin{array}{c}
7
\end{array}
\right)$,
$W^2_3 = 
\left(
\begin{array}{c}
7
\end{array}
\right)$,
$\al_3=10$.
\subsubsection{6.2.2 Recursive step. Triangular decomposition of matrix.}
$$\cB^2_4=\left(
\begin{array}{cc}
28&28\\
-7&-21\\
\end{array}
\right),\al_2=7$$
$$
\left\lbrace
P^4_6,
\bar L^4_6,
\lbrace \bar\al_4,\bar\al_5\rbrace,
\bar U^4_6,
Q^4_6,
\bar M^4_6,
\bar W^4_6
\right\rbrace
=\mathbf{LDU}(\cB^2_4,\al_2),$$
where  
$P^4_6 = 
\left(
\begin{array}{cc}
1&0\\
0&1\\
\end{array}
\right)$,
$\bar L^4_6 = \left(
\begin{array}{cc}
28&0\\
-7&-56\\
\end{array}
\right)=
\left(
\begin{array}{cc}
10&0\\
10&1\\
\end{array}
\right)$,
$\bar U^4_6=\left(
\begin{array}{cc}
28& 28\\
0&-56\\
\end{array}
\right)$,
$Q^4_6 = 
\left(
\begin{array}{cc}
1&0\\
0&1\\
\end{array}
\right)$,
$\bar M^4_6 = 
\left(
\begin{array}{cc}
7&0\\
7&28
\end{array}
\right)$,
$\bar W^4_6 = 
\left(
\begin{array}{cc}
7&-28\\
0&28
\end{array}
\right)$,
$\bar \al_4=28,\bar \al_5=-56$.
\subsubsection{Calculations before next recursive step.}
\begin{align}
L^4_6 & = \frac{\al_3}{\al_2} \bar L^4_6 = \frac{10}{7} \left(
\begin{array}{cc}
28&0\\-7&-56\\
\end{array}
\right)
=
\left(
\begin{array}{cc}
40&0\\
-10&-80\\
\end{array}
\right),
\nonumber\\
U^4_6 & = \frac{\al_3}{\al_2} \bar U^4_6 = \frac{10}{7} \left(
\begin{array}{cc}
28& 28\\
0&-56\\
\end{array}
\right)
=
\left(
\begin{array}{cc}
40&40\\
0&-80\\
\end{array}
\right),
\nonumber\\
M^4_6 & = \frac{\al_3}{\al_2} \bar M^4_6 = \frac{10}{7} 
\left(
\begin{array}{cc}
28&0\\
-7&-56\\
\end{array}
\right)
=
\left(
\begin{array}{cc}
10&0\\
10&40
\end{array}
\right),
\nonumber\\
W^4_6 & = \frac{\al_3}{\al_2} \bar W^4_6 = \frac{10}{7}\left(
\begin{array}{cc}
7&-28\\
0&28
\end{array}
\right)
=
\left(
\begin{array}{cc}
10&-40\\
0&40\\
\end{array}
\right),
\end{align}
 $
\al_4   = \frac{\al_3 \bar \al_4}{\al_2} = \frac{10\times28}{7} = 40, \ 
\al_5   = \frac{\al_3 \bar  \al_5}{\al_2} = \frac{10\times(-56)}{7} = -80.  
$

\noindent
Since $\cC^2_4$ has no full rank, we should split into blocks matrix $\cD^2_4$.
$P^2_4 \cD^2_4 Q^4_6= \left(
\begin{array}{c}
D_1\\
D_2
\end{array}
\right)$, where 
$D_1=
\left(
\begin{array}{cc}
12 &15
\end{array}
\right)$ and
$D_2 =
\left(
\begin{array}{cc}
12& 15
\end{array}
\right)$.

$$
U^d  = (\al_2)^{-1} M^2_3 D_1 = (7)^{-1} \left(
\begin{array}{c}
7
\end{array}
\right) \left(
\begin{array}{cc}
12& 15
\end{array}
\right)= 
\left(
\begin{array}{cc}
12& 15
\end{array}
\right),
$$ $$
L^d   = (\al_2)^{-1} (D_2 - M_0 (\frac{1}{\al_3}) U^d) W^4_6 = 
$$
$$
(7)^{-1} \left( \left(
\begin{array}{cc}
12& 15
\end{array}
\right) - \left(
\begin{array}{c}
10
\end{array}
\right)
(\frac{1}{10})\left(
\begin{array}{cc}
12& 15
\end{array}
\right) \right)
\left(
\begin{array}{cc}
10&-40\\
0&40\\
\end{array}
\right)=
\left(
\begin{array}{cc}
0&0
\end{array}
\right),
$$
$$
M^{bc} = - (\al_2)^{-1} M^4_6 0 (\frac{1}{\al_3}) M^2_3= 0, $$ $$
W^{bc} = - \frac{1}{\al_2}  W^2_3 \frac{1}{\al_3} U^d W^4_6 = 
-\frac{7}7   
\frac{1}{10}
  \left(
\begin{array}{cc}
12&15
\end{array}
\right)
\left(
\begin{array}{cc}
10&-40\\
0&40\\
\end{array}
\right)
= 
\left(
\begin{array}{cc}
-12&-12\\
\end{array}
\right).
\nonumber
$$
\subsubsection{6.2.3. Result of decomposition of matrix $\cA^2_6$}
$$\left\lbrace P^2_6,L^2_6,\lbrace\al_3,\al_4,\al_5\rbrace,U^2_6,Q^2_6,M^2_5,W^2_5 \right\rbrace = \mathbf{LDU}(\cA^2_6,\al_2),$$
where
$
P^2_6   = 
\left(
\begin{array}{cc}
0&I_2\\
I_2&0\\
\end{array}
\right)
\left(
\begin{array}{cc}
P^2_4&0\\
0&P^4_6\\
\end{array}
\right)
\left(
\begin{array}{ccc}
I_1 &0&0\\
0&0&I_1\\
0&I_2&0\\
\end{array}
\right)
,
Q^2_6 =
\left(
\begin{array}{ccc}
I_1 &0&0\\
0&0&I_2\\
0&I_1&0\\
\end{array}
\right)
\left(
\begin{array}{cc}
Q^2_4&0\\
0&Q^4_6\\
\end{array}
\right)
,\\
L^2_6   = 
\left(
\begin{array}{ccc}
L_0&0&0\\
0&L^4_6&0\\
M_0&L_d &I_1\\
\end{array}
\right)
,
U^2_6    =
\left(
\begin{array}{ccc}
U_0&U^d&V_0\\
0&U^4_6&0\\
0&0&I_1
\end{array}
\right)
,
M^2_6  =
\left(
\begin{array}{cc}
M^2_3&0\\
M^{bc}&M^4_6\\
\end{array}
\right)
=
\left(
\begin{array}{ccc}
7&0&0\\
0&10&0\\
0&10&40\\
\end{array}
\right),\\
W^2_6  =
\left(
\begin{array}{cc}
W^2_3&W^{bc}\\
0&W^4_6\\
\end{array}
\right)
=
\left(
\begin{array}{ccc}
7&-12&-12\\
0&10&-40\\
0&0&40\\
\end{array}
\right),
$
$\al_3   = 10, \al_4=40, \al_5 = -80, \\
D^2_6   = \diag(  
 \frac{1}{ \al_3},
 \frac{\al_2}{\al_3\al_4},
 \frac{\al_2}{\al_4\al_5},
 0 ), \
(P^2_6,  L^2_6, D^2_6, U^2_6, Q^2_6)  =$

\begin{small}
\noindent
$
\left(
\left[
\begin{array}{cccc}
0&1&0&0\\
0&0&1&0\\
1&0&0&0\\
0&0&0&1\
\end{array}
\right],
\left[
\begin{array}{cccc}
10& 0&   0  &0\\
0&  40&  0  &0\\
0& -10& -80&0\\
10 & 0 & 0 & 1\\
\end{array}
\right],
\left[
\begin{array}{cccc}
\frac{1}{10}&0&0&0\\
0&\frac{7}{400}&0&0\\
0&0&\frac{-7}{3200}&0\\
0&0&0&0\\
\end{array}
\right],
\left[
\begin{array}{cccc}
10& 12&   15  &-9\\
0&  40&  40  &0\\
0& 0& -80&0\\
0& 0 & 0 & 1\\
\end{array}
\right],
\left[
\begin{array}{cccc}
1&0&0&0\\
0&0&1&0\\
0&0&0&1\\
0&1&0&0\\
\end{array}
\right]\right)
$
\end{small}
\subsection{Final result.}
$$\left\lbrace P^0_6,L^0_6,\lbrace\al_1,\al_2\al_3,\al_4,\al_5\rbrace,U^0_6,Q^0_6\right\rbrace = \mathbf{LDU}(\cA^0_6,\al_0),$$ 
$$
D^0_6   = \diag( 
\frac{\al_0}{\al_0\al_1},
 \frac{\al_0}{\al_1\al_2}, 
 \frac{\al_0}{\al_2\al_3},
 \frac{\al_0}{\al_3\al_4},
 \frac{\al_0}{\al_4\al_5},
 0 )
  = \diag( 
\frac{1}{3},
 \frac{1}{21},
 \frac{1}{70},
 \frac{1}{400},
 -\frac{1}{3200},
0)
$$
$$
P^0_6   = 
\left(
\begin{array}{cc}
P^0_4&0\\
0&I_2\\
\end{array}
\right)
\left(
\begin{array}{cc}
I_2&0\\
0&P^2_6\\
\end{array}
\right)
, 
L^0_6   = 
\left(
\begin{array}{cc}
L^0_2&0\\
{P^2_6}^T\widetilde{L^0_2}&L^2_6\\
\end{array}
\right)
, 
U^0_6   = 
\left(
\begin{array}{cc}
U^0_2&\widetilde{U^0_2}{Q^2_6}^T\\
0&U^2_6\\
\end{array}
\right),
$$

\noindent
$$
Q^0_6   = 
\left(
\begin{array}{cc}
I_2&0\\
0&Q^2_6\\
\end{array}
\right)
\left(
\begin{array}{cc}
Q^0_4&0\\
0&I_2\\
\end{array}
\right),
(P^0_6, L^0_6, U^0_6, Q^0_6)=$$ 

\noindent
$
\left(\left[
\begin{array}{cccccc}
1&0&0&0&0&0\\
0&1&0&0&0&0\\
0&0&0&1&0&0\\
0&0&0&0&1&0\\
0&0&1&0&0&0\\
0&0&0&0&0&1\\
\end{array}
\right],
\left[
\begin{array}{cccccc}
3&0&0&0&0&0\\
1&7&0&0&0&0\\
2&-1&10& 0&   0  &0\\
3&0&0&  40&  0  &0\\
1&7&0& -10& -80&0\\
2&-1&10 & 0 & 0 & 1\\
\end{array}
\right],
\left[
\begin{array}{cccccc}
3& 2& 3&  1&  2&   5 \\
0& 7& 9&  8&  10&  1 \\
0& 0& 10& 12& 15&  -9\\
0& 0& 0&  40& 40&  0 \\
0& 0& 0&  0&  -80& 0 \\
0& 0& 0&  0&  0&   1 \\
\end{array}
\right],
\left[
\begin{array}{cccccc}
1&0&0&0&0&0\\
0&1&0&0&0&0\\
0&0&1&0&0&0\\
0&0&0&0&1&0\\
0&0&0&0&0&1\\
0&0&0&1&0&0\\
\end{array}
\right] 
\right)
$
$$
\cL = P^0_6 L^0_6 {P^0_6}^T, \
\cD = P^0_6 D^0_6 Q^0_6,\ 
\cU = {Q^0_6}^T U^0_6 Q^0_6.
$$
\section{Conclusion}
We offer an algorithm for the matrix  factorization, which has the complexity of matrix multiplication. It was 
  implemented at http://mathpartner.com.  
An example can be obtain as follows: ``A=[[1,2],[3,4]]; $\backslash$LDU(A)``.



\begin{thebibliography}{99}
%
\bibitem{08}
Grigoriev, D.: Analogy of Bruhat decomposition for the closure of
a cone of Chevalley group of a classical series. Soviet Math.
Dokl. 23, 393--397 (1981)
%
\bibitem{09}
Grigoriev, D.: Additive complexity in directed computations.
Theoretical Computer Science 19, 39--67 (1982)
%
\bibitem{10}
Akritas A.G., Akritas E.K., Malaschonok G.I.: Various Proofs of Sylvester’s (De-
terminant) Identity. In: Proceedings International IMACS Symposium on Symbolic
Computation, 14–17 June 1993, pp. 228--230. Lille, France (1993)
%
\bibitem{11}
Malaschonok, G.I.: Effective matrix methods in commutative domains. 
In: D. Krob, A.A. Mikhalev, A.V. Mikhalev (eds.) 
Formal Power Series and Algebraic 
Combinatorics, pp. 506--517. Springer, Berlin (2000)
%
\bibitem{12}
Malaschonok, G.I.: Matrix Computational Methods in Commutative
Rings. Tambov University Publishing House, Tambov (2002)
%
\bibitem{13}
Malaschonok, G.I.: Effective matrix methods in commutative
domains. In:  D. Krob, A.A. Mikhalev, A.V. Mikhalev (eds.) Formal
Power Series and Algebraic Combinatorics,  pp.~506--517. Springer,
Berlin  (2000)
%
\bibitem{14}
 Malaschonok, G.I.: A fast algorithm for adjoint matrix computation. 
 Tambov University Reports 5(1),  142--146  (2000)
%
\bibitem{15}
  Malaschon5k, G.I.: Fast matrix decomposition in parallel computer algebra. 
  Tambov University Reports 15(4),  1372--1385 (2010)
%
  \bibitem{16}
Malaschonok, G.I.: Fast generalized Bruhat decomposition. In:
Ganzha, V.M., Mayr, E.W., Vorozhtsov, E.V. (eds.) 12th
International Workshop on Computer Algebra  in Scientific
Computing (CASC 2010), pp.  194--202. LNCS 6244. Springer, Berlin,
Heidelberg (2010)
%
\bibitem{17}
Malaschonok, G.I.: On the fast generalized Bruhat decomposition in
domains. Tambov University Reports 17(2), 544--550 (2012)
%
\bibitem{18}
Malaschonok, G.I.: Generalized Bruhat decomposition in commutative domains. In:
International Workshop on Computer Algebra  in Scientific
Computing (CASC 2013), pp.  231--242. LNCS 8136. Springer, Berlin,
Heidelberg (2013)
\end{thebibliography}
\end{document}